\documentclass[a4paper,11pt]{article}
\usepackage{amsfonts}
\usepackage{amsmath}
\usepackage{amssymb}
\usepackage{graphicx}
\usepackage{array}
\usepackage{booktabs}
\usepackage{multirow}
\usepackage{makecell}
\usepackage{float}
\pdfoutput=1 

\usepackage{mymacros}
\usepackage{chngcntr}
\usepackage{verbatim}
\usepackage{amsthm}
\usepackage{graphics}
\usepackage{xcolor}
\usepackage{mathrsfs}
\usepackage{bbold}
\usepackage{cases}
\usepackage{epsfig}
\usepackage{epstopdf}
\usepackage{hyperref}
\usepackage{amsfonts}
\usepackage{hyperref}
\usepackage{jheppub} 

\usepackage[T1]{fontenc} 

\hypersetup{
	colorlinks=true,
	linkcolor=blue,
	filecolor=blue,
	urlcolor=blue,
	citecolor=blue,
}

\title{\boldmath Chaos bound violation by spinning particles in Gauss-Bonnet-AdS black holes}

\author{Xiaowei Li $^{a}$,}
\author{Mengxuan Wu $^{a}$ and}
\author{Guoping Li $^{b}$$\footnote{\texttt{Corresponding author:gpliphys@yeah.net}}$}

\affiliation{$^{a}$ School of Mathematics, Physics and Statistics, Sichuan Minzu College, Kangding 626001, People's Republic of China\\
$^{b}$  School of Physics and Astronomy, China West Normal University, Nanchong 637000, People's Republic of China}

\abstract{In this work, we investigate the violation of the chaos bound for spinning test particles in Gauss-Bonnet anti-de Sitter spacetime, focusing on the regulatory roles of the Gauss-Bonnet parameter and spacetime dimensionality. In five-dimensional spacetime, the Lyapunov exponent grows monotonically with particle spin. In contrast, in eight- and nine-dimensional spacetimes, it exhibits non-monotonic behavior-first decreasing and then increasing-reflecting the nonlinear nature of higher-dimensional tensor couplings. The Gauss-Bonnet parameter significantly modulates the violation by reshaping the near-horizon geometry: in five dimensions, the deviation of the Lyapunov exponent from the surface gravity first increases and then decreases with the Gauss-Bonnet parameter, whereas in the higher dimensions the bound is more easily violated. Increasing the total angular momentum of the particle also enhances chaos; however, when the Gauss-Bonnet parameter and black hole charge are small, the spacetime dimensionality, rather than the angular momentum, dominates. These results establish the spacetime dimensionality and the Gauss-Bonnet parameter as important factors governing the validity of the chaos bound in modified gravity theories.}

\begin{document}
	\maketitle
	\flushbottom
	
\section{Introduction}

Recent advances in holographic duality have motivated extensive investigations into chaotic dynamics near black hole horizons \cite{Maldacena:1997re}. In 2016, Maldacena, Shenker, and Stanford conjectured a universal upper bound (chaos bound) on the Lyapunov exponent (LE) in thermal quantum systems \cite{Maldacena:2015waa},

\begin{eqnarray}
\lambda\leq \frac{2\pi k_B T}{\hbar} , \label{eq1.1}
\end{eqnarray}

\noindent where $k_B$ is the Boltzmann constant and $T$ is the temperature of the system. This bound is saturated in the gravitational dual by the shockwave geometry near the black hole horizon \cite{Shenker:2013pqa}, providing an important clue to the deep connection between black hole thermodynamics and quantum chaos. This conjecture has been verified in the SYK model \cite{SYK1,SYK2,SYK3,SYK4,SYK5}, where the bound is saturated.

The classical test of this conjecture was initiated by Hashimoto and Tanahashi \cite{Hashimoto}, who examined the chaotic dynamics of test particles near the horizon of a spherically symmetric black hole by computing the LE. They found that the exponent depends only on the surface gravity $\kappa$  of the black hole and is independent of the particle species and external forces, satisfying

\begin{eqnarray}
\lambda\leq \kappa. \label{eq1.2}
\end{eqnarray}

\noindent This result reveals the universality of chaotic dynamics near the horizon and, since the black hole temperature is proportional to the surface gravity, formally aligns the classical bound with the quantum one, lending support to the conjecture at the classical level.

Subsequently, Zhao et al. incorporated the Lorentz force acting on the particle and analytically derived the LE by analyzing the particle effective potential under static equilibrium near the black holes, then quantitatively compared it with the saturation value of the chaos bound. Their results showed that violations of the chaos bound can occur for particles around most black holes, with the exception of Reissner-Nordström (RN) and Reissner-Nordstr\"{o}m anti-de Sitter (RN-AdS) black holes \cite{ZLL}. This work uncovered the cooperative regulation of the chaos bound by spacetime geometry and electromagnetic interactions, triggering extensive tests of the bound in various gravitational backgrounds. Within the framework of general relativity (GR), the chaos bound has been examined in RN \cite{LG1,LG4}, Kerr-Newman \cite{KG1}, Kerr-Newman-(A)dS \cite{KG2,KG3}, Kiselev \cite{GC1,LMX1}, and Taub-NUT black holes \cite{YC1}, among others \cite{LG2,SP1,SP2,SP3,SYHM1,SYHM2,SYHM3,SYHM4,SYHM5,SP4,SP5}. The results consistently show that the bound is violated under certain conditions, with the exponent deviating more strongly from the surface gravity as the black hole approaches extremality. These tests have also been extended to broader gravitational theories, including Einstein-Maxwell-Dilaton-Axion theory \cite{KG4,HLBG,YCG1}, supergravity \cite{LG3,SSS1}, and Einstein-Euler-Heisenberg gravity  \cite{GC2}.

From a theoretical standpoint, GR is a low-energy effective theory that inevitably receives higher-curvature corrections in high-curvature regimes. Gauss-Bonnet gravity, as the second-order Lovelock theory, is one of the most natural extensions beyond Einstein gravity. The Gauss-Bonnet term is a topological invariant in four dimensions but yields nontrivial dynamics in five dimensions and above. Crucially, it arises naturally in the low-energy effective action of string theory, making Gauss-Bonnet-AdS spacetime an important theoretical arena for testing holographic duality and the chaos bound. In Gauss-Bonnet gravity, higher-curvature corrections modify the spacetime geometry and consequently affect test-particle dynamics. Existing studies have shown that the Gauss-Bonnet coupling significantly influences black hole stability, quasinormal modes, and thermodynamic properties. A natural question then arises: how does this coupling affect the chaos bound? Although some works have addressed this question for scalar particles in Gauss-Bonnet gravity, they are restricted to four-dimensional cases \cite{FA1,FA2}. Higher-dimensional gravity is more meaningful in this context, since the Gauss-Bonnet term is a topological invariant in four dimensions, where its field equations reduce to the standard Einstein equations and produce no nontrivial dynamical corrections. Only in five dimensions and above does the Gauss-Bonnet term exert a genuine impact on spacetime geometry and particle motion, thereby revealing the physical effects of higher-curvature corrections on the chaos bound.

Meanwhile, a particle spin, as an intrinsic property of test particles, plays an indispensable role in chaos bound studies. Most previous works have adopted spinless scalar particles, neglecting spin-orbit coupling. In reality, however, particles generically carry spin, and the coupling between spin and spacetime curvature causes the four-velocity and four-momentum to deviate from parallelism, thereby altering the particle's trajectory and LE. In recent work, spinning particles have been introduced in spherically symmetric spacetimes to test the chaos bound, but these studies often lack rigorous physical constraints \cite{YCL1,YCL2}, leaving some conclusions potentially contaminated by unphysical configurations. A systematic investigation of spinning-particle chaos in Gauss-Bonnet-AdS spacetime under physically reasonable parameter ranges is therefore a meaningful undertaking.

In this work, we systematically test the chaos bound using spinning test particles in higher-dimensional, spherically symmetric Gauss-Bonnet-AdS spacetime. We adopt the Hamiltonian formalism for spinning particles developed by Hojman, Asenjo, and others to derive the equations of motion, and then compute the LE near equilibrium orbits, comparing it with the saturation value predicted by the chaos bound. Throughout the analysis, we impose strict physical constraints to ensure that the particle motion remains within physically admissible regions, thereby excluding unphysical configurations from affecting our conclusions. Our study focuses on the regulatory roles of the Gauss-Bonnet coupling constant, the particle spin parameter, and the cosmological constant in the violation of the chaos bound. Through spin-orbit coupling, the particle spin introduces directional dependence, rendering the conditions for bound violation more intricate. The negative cosmological constant acts as a potential well: the larger its magnitude, the more pronounced the chaotic behavior near the horizon. When discussing the effects of the Gauss-Bonnet coupling, particle spin, and spacetime dimensionality on the chaos bound, we fix the AdS radius, because a fixed radius ensures that all dimensions share the same AdS curvature scale. If the cosmological constant were fixed instead, the AdS radius would necessarily change with dimensionality, meaning that black holes in different dimensions would reside in spacetimes with entirely different curvature scales. In that case, any observed differences in chaotic behavior could not be unambiguously attributed to dimensionality alone, as they might instead reflect the altered background curvature scale.

The rest is organized as follows. In Sec. \ref{sec2}, we briefly review the thermodynamics of Gauss-Bonnet-AdS black holes and the equations of motion for spinning particles, and present the expression for the LE. In Sec. \ref{sec3}, we numerically test the chaos bound for spinning test particles and discuss the dependence of the exponent on the black hole and particle parameters. Sec. \ref{sec4} is devoted to a summary and discussion.

\section{Dynamics of spinning particles in Gauss-Bonnet-AdS spacetime}\label{sec2}

\subsection{Thermodynamics of black holes}\label{sec2.1}

The $d$-dimensional Einstein-Maxwell theory with a Gauss-Bonnet term and a cosmological constant $\Lambda$ is described by the action

\begin{eqnarray}
S=\frac{1}{16\pi}\int d^{d}x\sqrt{-g}\left(R-2\Lambda+\alpha \mathcal{L}_{{GB}} -4\pi F_{\mu\nu}F^{\mu\nu}\right), \label{eq2.1.1}
\end{eqnarray}

\noindent with the Gauss-Bonnet Lagrangian $\mathcal{L}_{{GB}}= \mathcal{R}_{\mu\nu\gamma\delta} \mathcal{R}^{\mu\nu\gamma\delta}-4\mathcal{R}_{\mu\nu} \mathcal{R}^{\mu\nu}+\mathcal{R}^{2}$. $\alpha$ is the Gauss-Bonnet parameter and has dimension [length]$^2$, and the cosmological constant is related to the AdS radius $l$ via $\Lambda = -\frac{(d-1)(d-2)}{2l^2}$. The Maxwell field strength is defined by $F_{\mu\nu}=\partial _{\mu}A_{\nu}-\partial _{\nu}A_{\mu}$ with the electric potential $A_{\mu}$ . From the action, the solution for a spherically symmetric black hole is obtained as \cite{BD1,BD2,BD3,BD4}

\begin{eqnarray}
ds^2&=&-f(r)dt^2+\frac{1}{f(r)}dr^2+r^2d\Omega^2_{d-2}, \label{eq2.1.2}\\
f(r)&=&1+\frac{r^2}{2\tilde{\alpha}}\left(1-\sqrt{1+\frac{64\pi\tilde{\alpha} M}{(d-2)\omega_{d-2} r^{d-1}}-\frac{8\tilde{\alpha} Q^2}{(d-2)(d-3)r^{2d-4}}-\frac{64\pi\tilde{\alpha}P}{(d-1)(d-2)}}\right), \label{eq2.1.3}
\end{eqnarray}

\noindent where $M$ and $Q$ denote the mass and charge of the black hole, respectively. The parameter $\tilde{\alpha}=(d-3)(d-4)\alpha$ also has dimension [length]$^2$, and $\omega_{d-2}=2\pi^{(d-1)/2}/\Gamma((d-1)/2)$ is the volume of the unit $(d-2)$-sphere. The electromagnetic potential is $A_t=-\frac{\omega_{d-2}Q}{4\pi(d-3)r^{d-3}}$. Here the cosmological constant is interpreted as the pressure $P=-\frac{\Lambda}{8\pi}$, and its conjugate thermodynamic volume is

\begin{eqnarray}
V=\frac{\omega_{d-2}r_+^{d-1}}{d-1}.\label{eq2.1.4}
\end{eqnarray}

The surface gravity is

\begin{eqnarray}
\kappa =\frac{-Q^2 r_+^{7-2 d} + 16 \pi  P 	r_+^3 +(d-3) (d-2) r_+}{(d-2) (2 \alpha
	+r_+^2)}-\frac{5 \alpha -\alpha  d}{4\alpha  r_++ 2  r_+^3}. \label{eq2.1.5}
\end{eqnarray}

\noindent In the above equation, $r_+$ is the radius of the event horizon determined by $f(r_+) = 0$. The black hole temperature is related to the surface gravity via $T=\frac{\kappa}{2\pi}$. The entropy $S$, electric potential $\Phi$ at the event horizon, and conjugate quantity $\mathcal{A}$ to $\alpha$ are given by \cite{CCLY,WL}

\begin{eqnarray}
S&=&\frac{\omega_{d-2}r_+^{d-2}}{4}\left(1+\frac{2(d-2)\alpha}{(d-4)r_+^2}\right), \label{eq2.1.6}\\
\Phi&=&\frac{\omega_{d-2}Q}{4\pi(d-3)r_+^{d-3}}, \label{eq2.1.7}\\
\mathcal{A}&=&-\frac{\omega_{d-2}  r_+^{-d-5} \left(r_+^{2 d} \left(-2
	\alpha (d-2)+(d-2)^2 r_+^2+32 \pi P r_+^4\right)-4Q^2 r_+^8\right)}{16 \pi  (d-4)
	\left(2 \alpha +r_+^2\right)}. \label{eq2.1.8}
\end{eqnarray}\\

\noindent These thermodynamic quantities satisfy the first law of thermodynamics

\begin{eqnarray}
dM = TdS+\Phi dQ+\mathcal{A}d\alpha+VdP. \label{eq2.1.9}
\end{eqnarray}

\subsection{Equations of motion for spinning particles}\label{sec2.2}

In this work, we consider a spinning test particle of  mass $m$ and charge $\tilde{q}$ moving in this spacetime. Its dynamics are governed by the Mathisson-Papapetrou-Dixon (MPD) equations \cite{RH},

\begin{align}
\frac{Dp^\mu}{D\tau} &= -\frac12 R^\mu_{\nu\alpha\beta} u^\nu S^{\alpha\beta} -  \tilde{q} F^\mu_{\ \nu} u^\nu, \label{eq:mpd1}\\
\frac{DS^{\mu\nu}}{D\tau} &= p^\mu u^\nu - u^\mu p^\nu, \label{eq:mpd2}
\end{align}

\noindent where \(\frac{D}{D\tau}\) is the covariant derivative along the particle's worldline, parametrized by  \(\tau\). Here \(p^\mu\) and \(u^\mu = \frac{dx^\mu}{d\tau}\) denote the four-momentum and four-velocity, respectively, \(R^\mu_{\ \nu\alpha\beta}\) is the Riemann tensor, and \(S^{\mu\nu}\) is the antisymmetric spin tensor. To establish a well-defined relation between the four-velocity and the four-momentum of the particle, it is necessary to impose a spin supplementary condition. In this work, we adopt the Tulczyjew-Dixon spin supplementary condition \cite{RH,WT},

\begin{equation}
S^{\mu \nu}p_{\nu} = 0.\label{eq2.6}
\end{equation}

\noindent The conserved mass $m$ and spin magnitude \( \tilde{S} \) are then defined by

\begin{eqnarray}
m^2 &=& (p^t)^2 f - \dfrac{(p^r)^2}{f} - r^2 (p^\phi)^2, \label{eq2.7.1}\\
\tilde{S}^2 &=& m^2S^2= \dfrac{r^2 (S^{r\phi})^2}{f}-(S^{tr})^2 - f r^2 (S^{t\phi})^2. \label{eq2.7}
\end{eqnarray}

\noindent   We adopt the method developed in \cite{NZA} to solve the particle's equations of motion. Expanding the Tulczyjew-Dixon condition in the Gauss-Bonnet-AdS spacetime, we obtain

\begin{equation}
\begin{aligned}
\dfrac{p^r S^{tr}}{f} + r^2 p^\phi S^{t\phi} = 0, \label{eq2.9}
\end{aligned}
\end{equation}
\begin{equation}
\begin{aligned}
r^2 p^\phi S^{r\phi} + f p^t S^{tr} = 0,\label{eq2.10}
\end{aligned}
\end{equation}
\begin{equation}
\begin{aligned}
\dfrac{1}{f} p^r S^{r\phi} - f p^t S^{t\phi} = 0. \label{eq2.11}
\end{aligned}
\end{equation}

\noindent The momentum equations \eqref{eq:mpd1} take the explicit form

\begin{equation}
\begin{aligned}
\dot{p}^t +\frac{p^r f'}{2f} +\frac{\dot{r}p^t f'}{2f} -\frac{r\dot{\phi}S^{t\phi}f'}{2} -\frac{\dot{r}S^{tr}f'}{2} +\frac{\dot{r}\tilde{q}A_{t,r}}{f} &= 0, \label{eq2.12}
\end{aligned}
\end{equation}
\begin{equation}
\begin{aligned}
\dot{p}^r +\frac{p^t f f'}{2} -\dot{\phi} p^\phi f' r -\frac{r\dot{\phi}S^{r\phi}f'}{2} -\frac{S^{tr}f f'}{2} -\frac{\dot{r}p^r f'}{2f} +\frac{\dot{r}\tilde{q}A_{t,r}}{f} +\tilde{q}A_{t,r}f &= 0,\label{eq2.13}
\end{aligned}
\end{equation}
\begin{equation}
\begin{aligned}
\dot{p}^\phi +\frac{\dot{\phi}p^r}{r} +\frac{\dot{r}p^\phi}{r} -\frac{\dot{r}S^{r\phi}f'}{2fr} -\frac{S^{t\phi}f f'}{2r} &= 0. \label{eq2.14}
\end{aligned}
\end{equation}

\noindent Here, the dot denotes differentiation with respect to the coordinate time, and the prime denotes differentiation with respect to the radial coordinate. The spin equation follows from \eqref{eq:mpd2} and reads

\begin{equation}
\begin{aligned}
\dot{S}^{tr} + p^r - \dot{r} p^t - f r \dot{\phi} S^{t\phi} &= 0,\label{eq2.15}
\end{aligned}
\end{equation}
\begin{equation}
\begin{aligned}
\dot{S}^{t\phi} - \dot{\phi} p^t + p^\phi + \frac{\dot{\phi} S^{tr} + \dot{r} S^{t\phi}}{r} + \frac{S^{r\phi} f' + \dot{r} S^{t\phi} f'}{2f} &= 0, \label{eq2.16}
\end{aligned}
\end{equation}
\begin{equation}
\begin{aligned}
\dot{S}^{r\phi} - \dot{\phi} p^r + \dot{r} p^\phi + \frac{\dot{r} S^{r\phi}}{r} + \frac{S^{t\phi} f f'}{2} - \frac{\dot{r} S^{r\phi} f'}{2f} &= 0.\label{eq2.17}
\end{aligned}
\end{equation}

\noindent The geometry possesses two Killing vector fields, $\xi_{(t)}^a=\left(\frac{\partial}{\partial t}\right)^a$ and  $\xi_{(\phi)}^a=\left(\frac{\partial}{\partial \phi}\right)^a$, corresponding respectively to stationarity and axisymmetry.  From the timelike Killing vector $\xi_{(t)}^a$, we obtain a conserved quantity identified with the energy,

\begin{align}
\tilde{E} =  p^{t}f -\frac{1}{2}S^{tr} f'  -  \tilde{q} A_t, \label{eq2.18}
\end{align}

\noindent where the prime indicates differentiation with respect to the radial coordinate $r$. The axial Killing vector $\xi_{(\phi)}^a$ yields the conserved angular momentum along the $z$-axis,

\begin{align}
\tilde{L} = r^{2}p^{\phi}+ rS^{r\phi}.\label{eq2.19}
\end{align}

\noindent For the numerical analysis below, we introduce the specific quantities $E=\frac{\tilde{E}}{m}$, $L=\frac{\tilde{L}}{m}$ and $q=\frac{\tilde{q}}{m}$, denoting the energy, total angular momentum and charge per unit mass, respectively. We also introduce $S=\pm\frac{\bar{S}}{m}$ to represent the angular momentum per unit mass of the particle, with a positive sign indicating alignment of the particle spin with the $z$-axis, and a negative sign indicating anti-alignment. Solving  Eqs.~\eqref{eq2.9}-\eqref{eq2.18} yields the radial equation of motion

\begin{align}
\dot r = \frac{p^{r}}{p^{t}}, \label{eq2.20}
\end{align}

\noindent with

\begin{align}
p^{t} &= -\frac{2 r (E + qA_t) -  S L f'}{(2 r -  S^{2} f') f}, \\
p^{\phi} &= -\frac{L - (E + qA_t) S}{\frac{1}{2}r  S^{2} f' -r^{2}}, \label{eq2.21}\\
p^{r} &= \pm \sqrt{ p_t^{2} - f\!\left(1+ \frac{p_\phi^{2}}{r^{2}}\right) }. \label{eq2.22}
\end{align}

\noindent The $\pm$ sign in Eq.~\eqref{eq2.22} distinguishes outgoing and incoming trajectories. We omit the $\phi$-direction equation of motion for brevity, as our focus is on the radial dynamics.

\subsection{Derivation of Lyapunov exponents}\label{sec2.3}

For a scalar particle, the LE that characterizes chaotic motion is governed by the effective potential and can be expressed in terms of its second derivative \cite{CMBWZ}. In contrast, a spinning particle possesses an enlarged phase space, and its dynamics are not reducible to those of a scalar particle. The corresponding exponent for a spinning particle has been derived in Refs. \cite{SSS4,YCL1,CHL}. In this section, we briefly revisit the derivation of this exponent for chaotic motion in the spinning particle, starting from the equations of motion.

\begin{align}
\frac{1}{2}m\dot{r}^2 +\mathcal{V}_{eff}=0,
\label{eq2.3.1}
\end{align}

\noindent where $\mathcal{V}_{eff}$ is the effective potential of the spinning particle.
From Eqs.~\eqref{eq2.3.1} and~\eqref{eq2.20}, we  arrive at its expression,

 \begin{equation}
\begin{aligned}
V_{\text{eff}}(r_0)=-\frac{m}{2}\left(\frac{p^r}{p^t}\right)^2\bigg|_{r = r_0}. \label{eq3.2.4}
\end{aligned}
\end{equation}

\begin{figure}[h]
	\begin{minipage}[t]{0.8\textwidth}
		\centering
		\includegraphics[width=10cm,height=8cm]{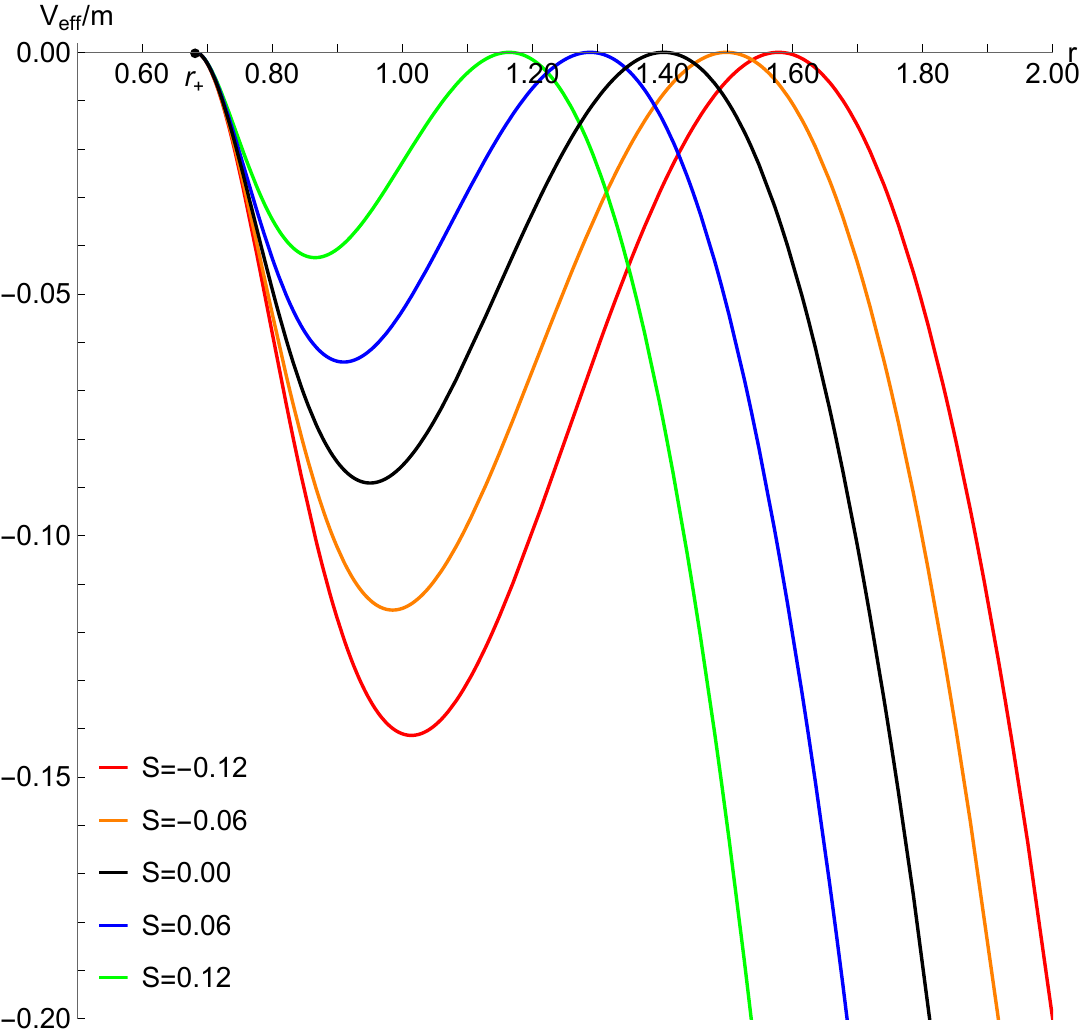}
	\end{minipage}
	\caption{Curves show the effective potential of the particle as a function of the radius, with $d=5$, $l^2=1.00$, $\alpha=0.04$, $Q=0.70$, $q=0.10$ and $L=10.00$. }
	\label{f1}
\end{figure}

\noindent Figure \ref{f1} shows that the extrema of the effective potential depend on both the magnitude and the direction of the particle spin. When the spin is aligned with the $z$-axis and takes its maximum value, the extremum of the effective potential lies closest to the black hole horizon. As the spin  increases in the opposite direction, the extremum gradually moves away from the horizon, reaching its farthest position at $S=-0.12$. The spinless case falls between these two extremes. The physical origin of this behavior lies in the spin-curvature coupling. The modulation of the extremum position by the spin essentially reveals the interaction between the spin and the spacetime geometry. The direction and magnitude of the spin alter the effective geometry experienced by the particle, biasing the curvature field it perceives. This implies that the spin breaks the spherical symmetry of the spacetime response, rendering the gravitational binding dependent on the intrinsic properties of the particle.

We focus on the regime where the particle is near an unstable equilibrium orbit at radius $r_0$. Writing the radial coordinate as $r(t)= r_0 +\epsilon(t)$ and expanding Eq. (\ref{eq2.3.1}) perturbatively in the small fluctuation $\epsilon$ around $r_0$, we obtain

\begin{eqnarray}
\frac{1}{2}\left(m\dot{\epsilon}^2 + \mathcal{V}^{\prime\prime}_{eff}(r_0) \epsilon^2\right) + \mathcal{V}_{eff}(r_0)+ \mathcal{O}(\epsilon) \simeq 0,
\label{eq2.3.2}
\end{eqnarray}

\noindent where $\mathcal{O}(\epsilon)$ collects terms of higher order in  $\epsilon$. We adopt the approach in \cite{CMBWZ} to determine the potential $\mathcal{V}_{eff}(r_0)=0$. Neglecting the higher-order contributions, the solution takes the exponential form $r(t)\approx r_0 + \epsilon(0)e^{\lambda t}$ with the LE obtained as

\begin{eqnarray}
\lambda^2 =-\frac{\mathcal{V}^{\prime\prime}_{eff}(r_0)}{m}=  \frac{1}{2}\frac{d^2}{dr^2}\left(\frac{p^r}{p^t}\right)^2\bigg|_{r = r_0}.
\label{eq2.3.3}
\end{eqnarray}

\noindent A positive LE indicates that the particle is in a chaotic state. If the exponent exceeds the surface gravity, this signals a violation of the chaos bound.

\section{Chaos bound for spinning particles: A numerical test}\label{sec3}

\begin{figure}[h]
	\centering
	\begin{minipage}[t]{0.48\textwidth}
		\centering
		\includegraphics[width=7cm,height=6cm]{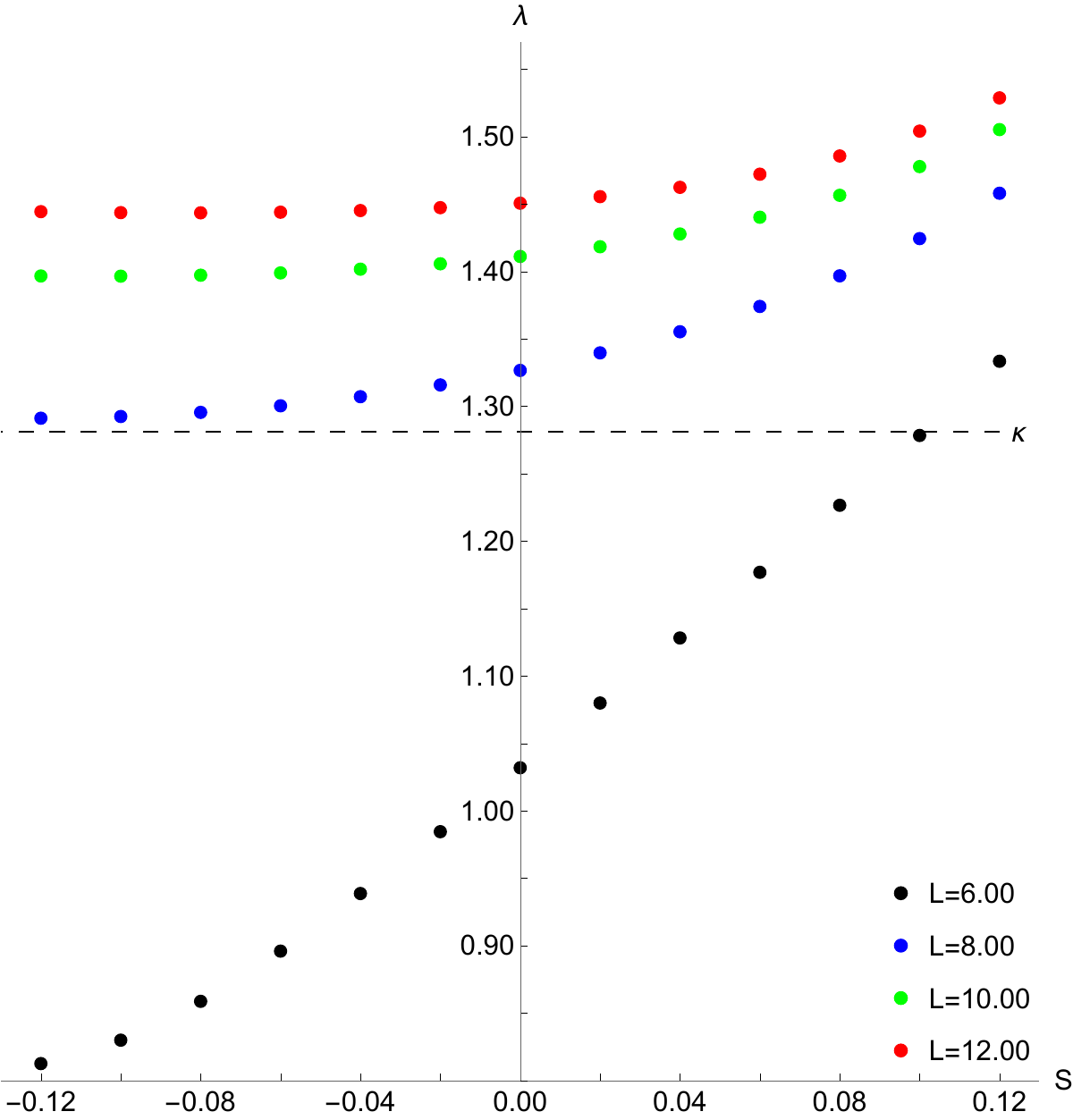}
		\subcaption{$\alpha=0.04$, $Q=0.70$, $d=5$.}
		\label{f21}
	\end{minipage}
	\begin{minipage}[t]{0.48\textwidth}
		\centering
		\includegraphics[width=7cm,height=6cm]{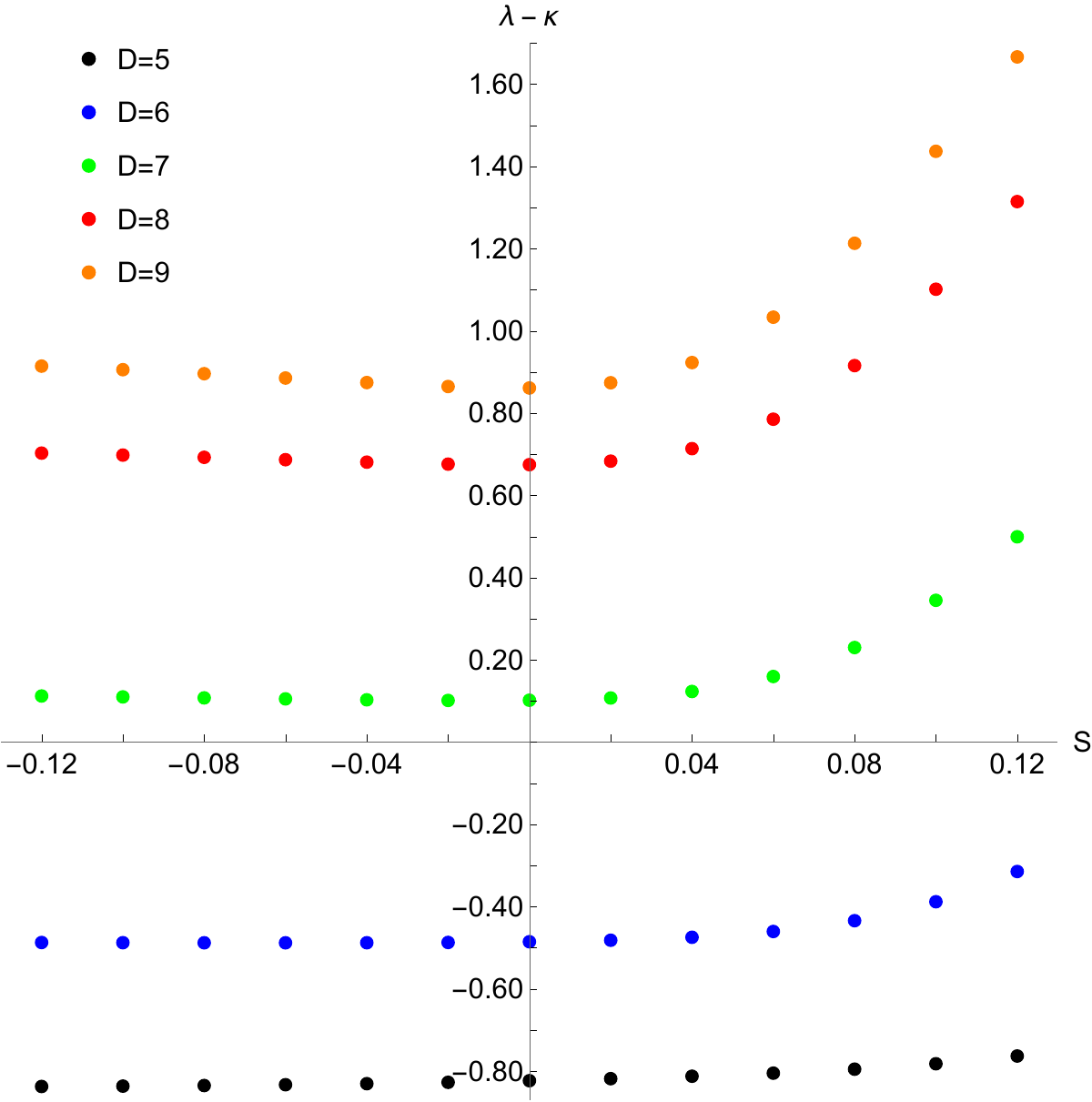}
		\subcaption{ $\alpha=0.005$, $Q=0.50$, $L=10.00$.}
		\label{f22}
	\end{minipage}
	\caption{Violation of the chaos bound as a function of the particle spin. Panel (a) presents the five-dimensional case, while panel (b) shows the results for spacetimes of different dimensions.}
\label{f2}
\end{figure}

\begin{figure}[h]
	\centering
	\begin{minipage}[t]{0.48\textwidth}
		\centering
		\includegraphics[width=7cm,height=6cm]{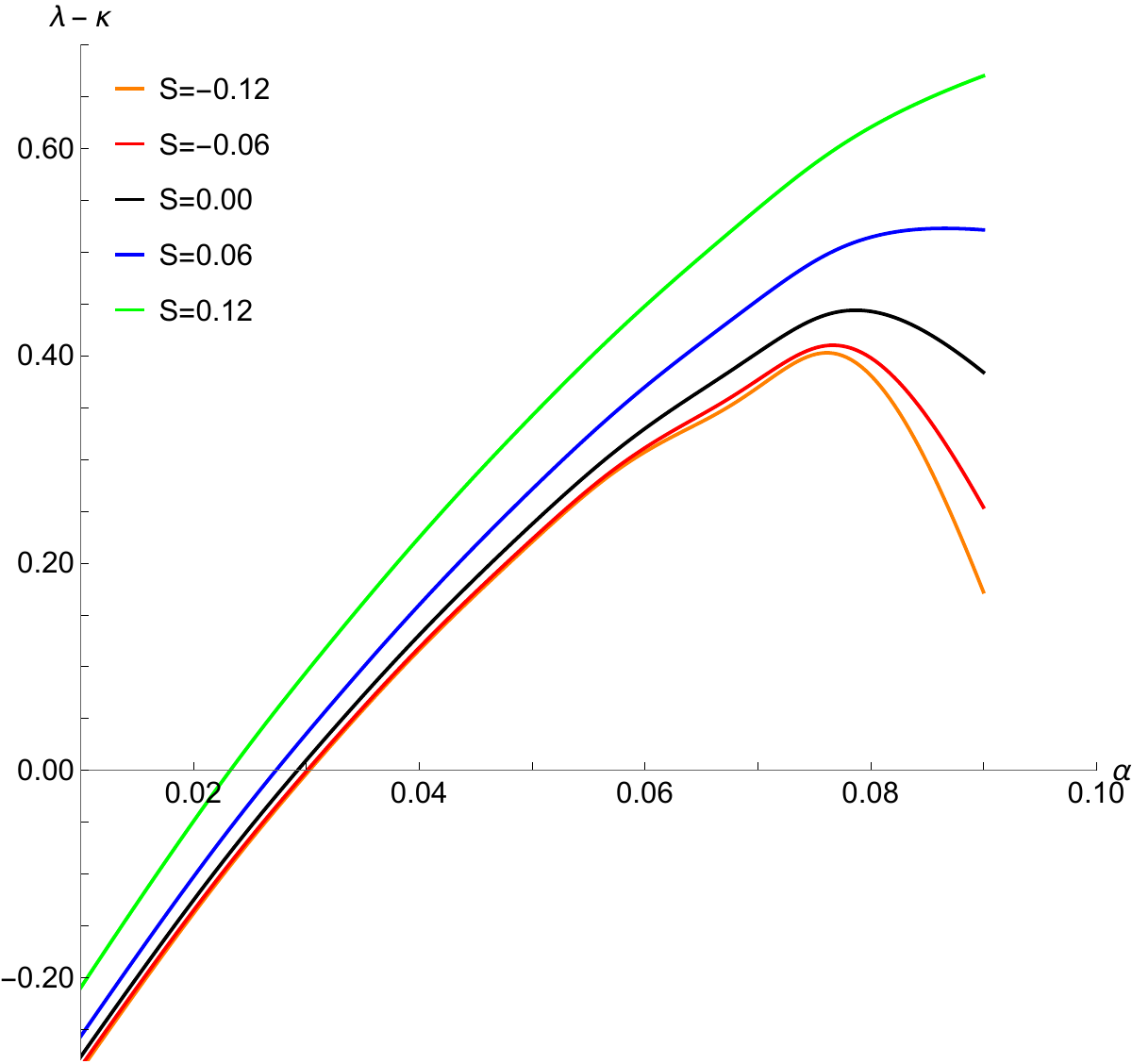}
		\subcaption{$Q=0.70$, $d=5$.}
		\label{f31}
	\end{minipage}
	\begin{minipage}[t]{0.48\textwidth}
		\centering
		\includegraphics[width=7cm,height=6cm]{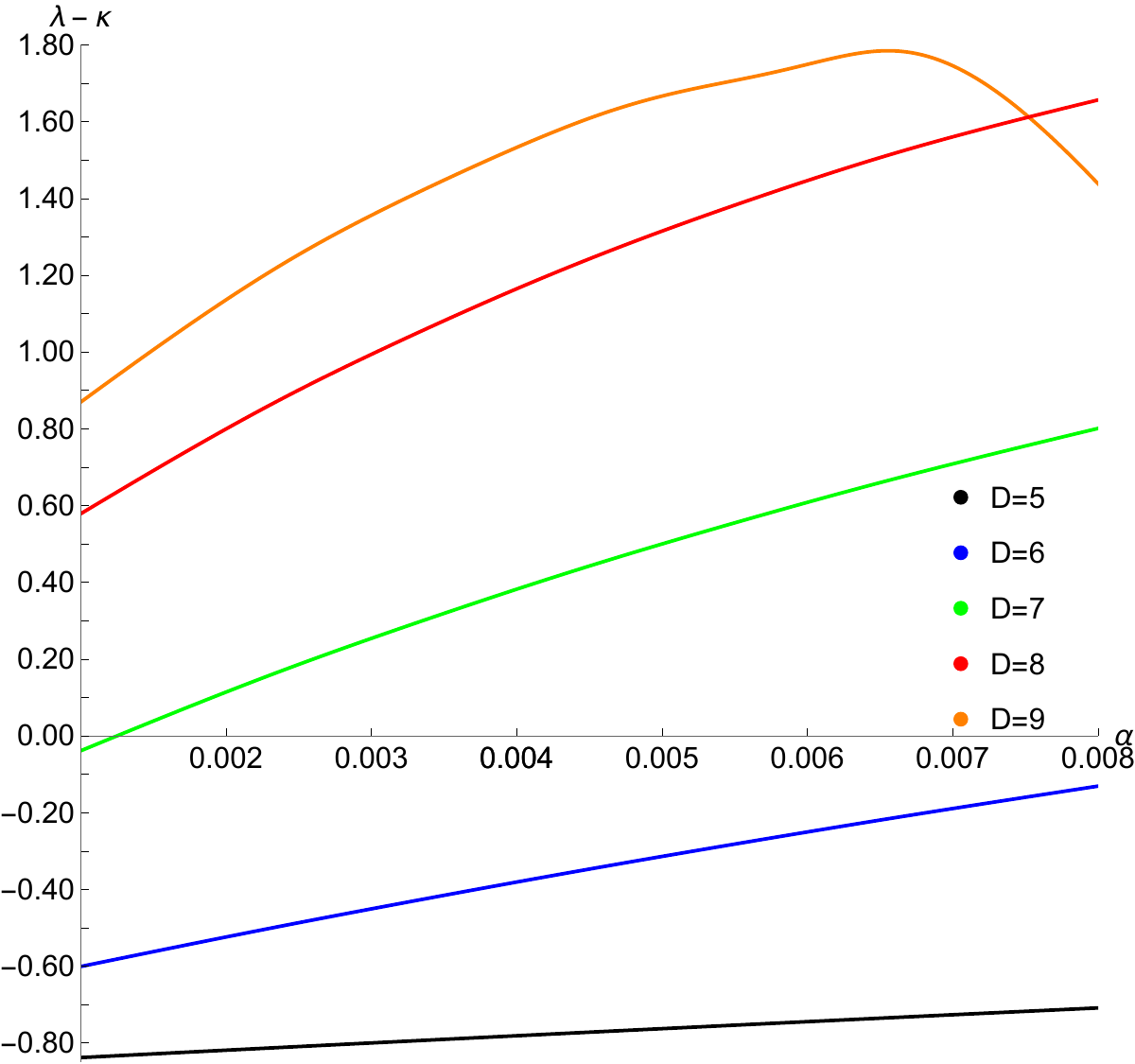}
		\subcaption{$Q=0.50$, $S=0.12$.}
		\label{f32}
	\end{minipage}
	\caption{Violation of the chaos bound as a function of the Gauss-Bonnet parameter. Panel (a) depicts the five-dimensional case, while panel (b) presents the results for spacetimes of various dimensions.}
	\label{f3}
\end{figure}

\begin{figure}[h]
	\centering
	\begin{minipage}[t]{0.48\textwidth}
		\centering
		\includegraphics[width=7cm,height=6cm]{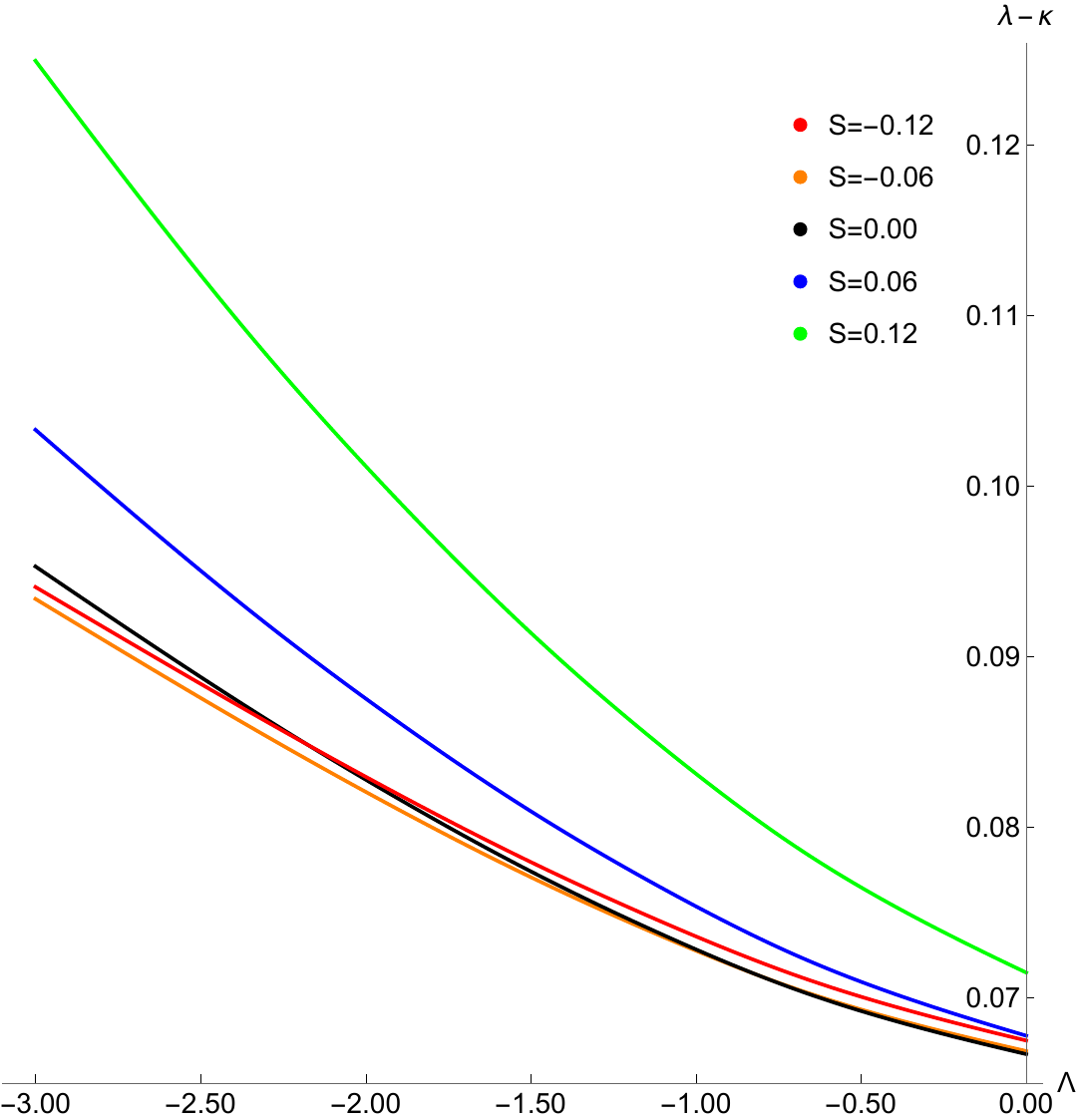}
		\subcaption{$\alpha=0.04$,  $Q=0.70$, $d=5$.}
		\label{f41}
	\end{minipage}
	\begin{minipage}[t]{0.48\textwidth}
		\centering
		\includegraphics[width=7cm,height=6cm]{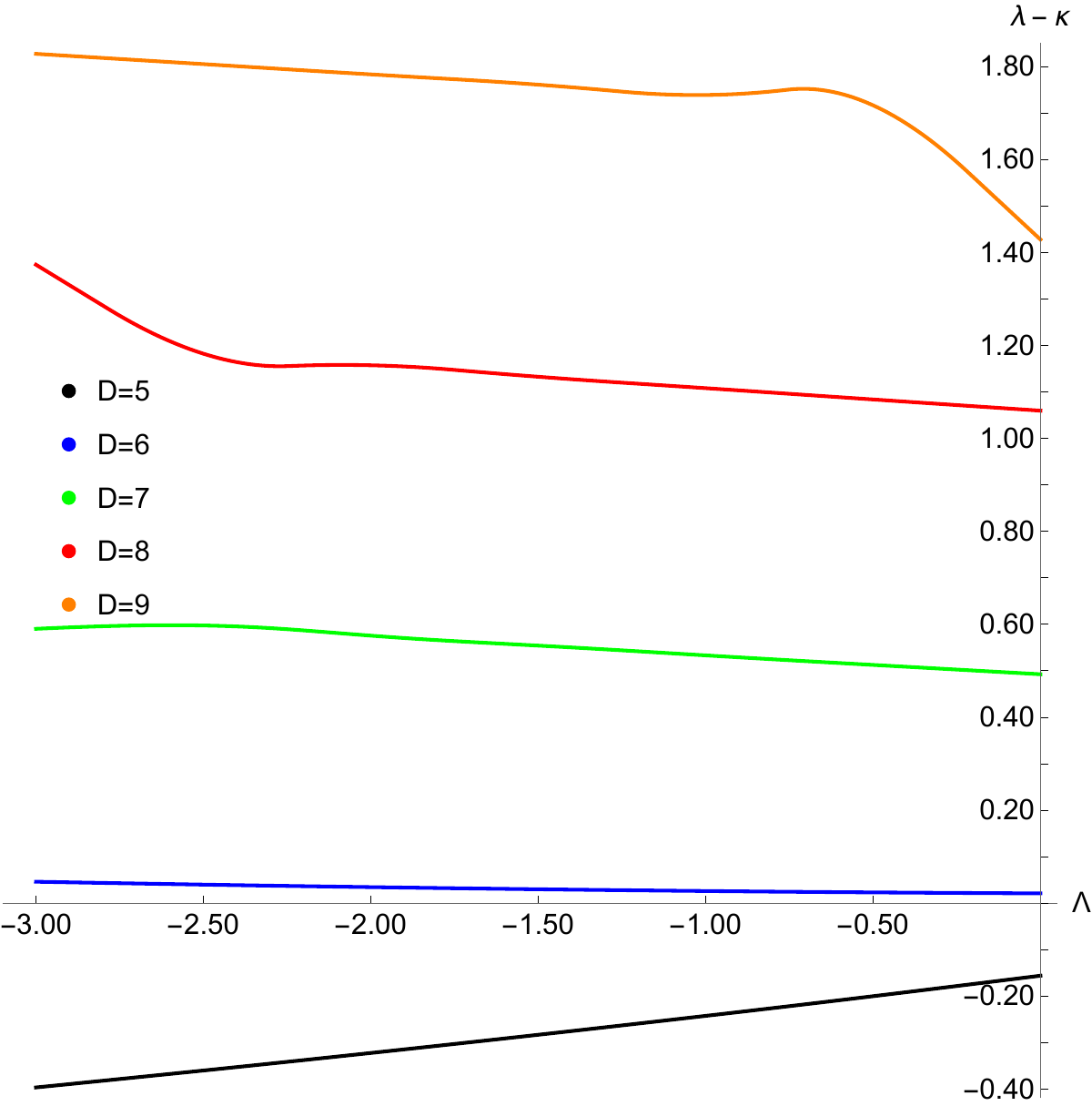}
		\subcaption{$\alpha=0.015$, $Q=0.50$,  $S=0.12$.}
		\label{f42}
	\end{minipage}
	\caption{Violation of the chaos bound as a function of the cosmological constant. Panel (a) depicts the five-dimensional case, while panel (b) presents the results for spacetimes of various dimensions.}
	\label{f4}
\end{figure}

\begin{figure}[h]
	\centering
	\begin{minipage}[t]{0.48\textwidth}
		\centering
		\includegraphics[width=7cm,height=6cm]{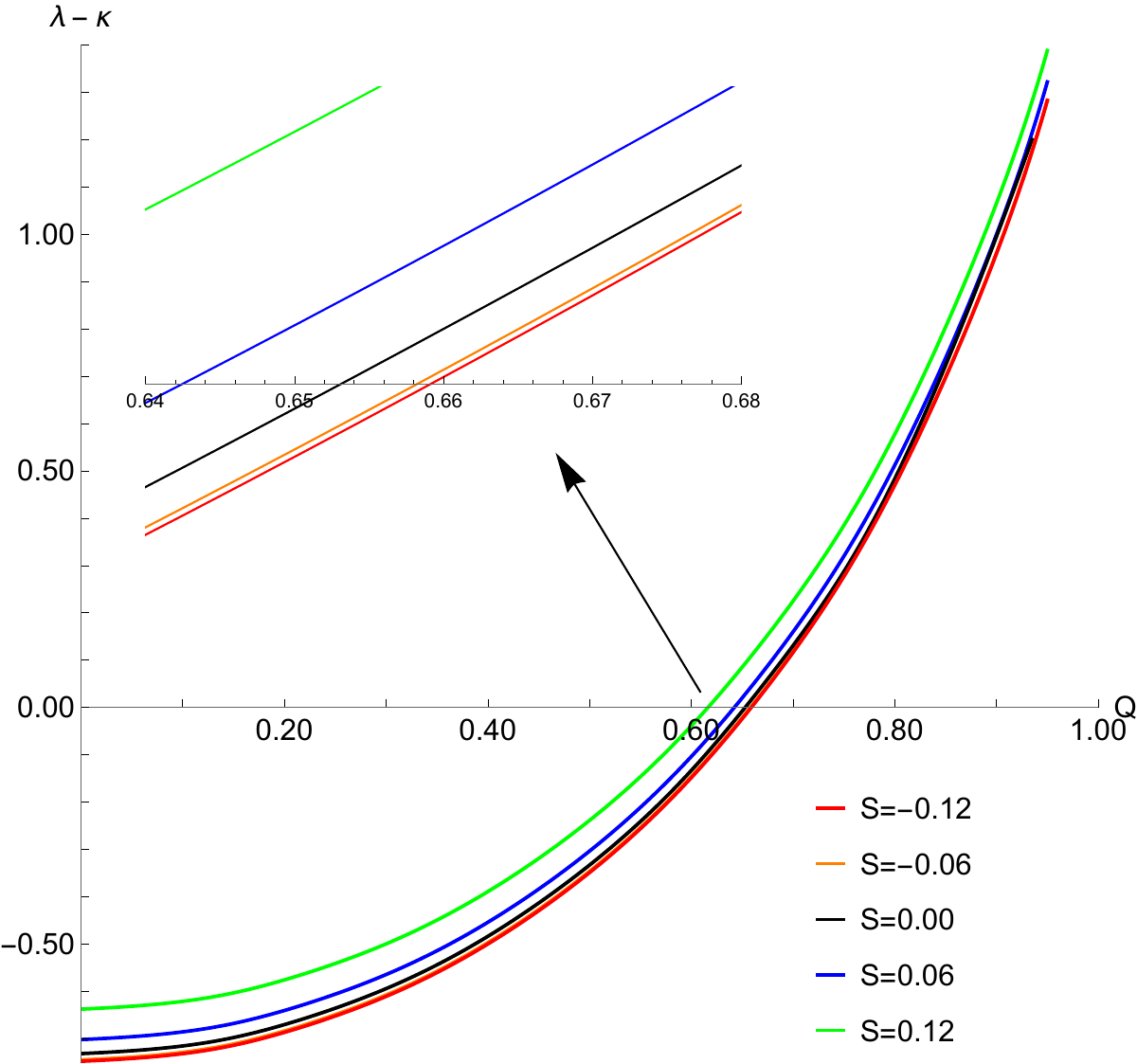}
		\subcaption{$\alpha=0.015$, $d=5$.}
		\label{f51}
	\end{minipage}
	\begin{minipage}[t]{0.48\textwidth}
		\centering
		\includegraphics[width=7cm,height=6cm]{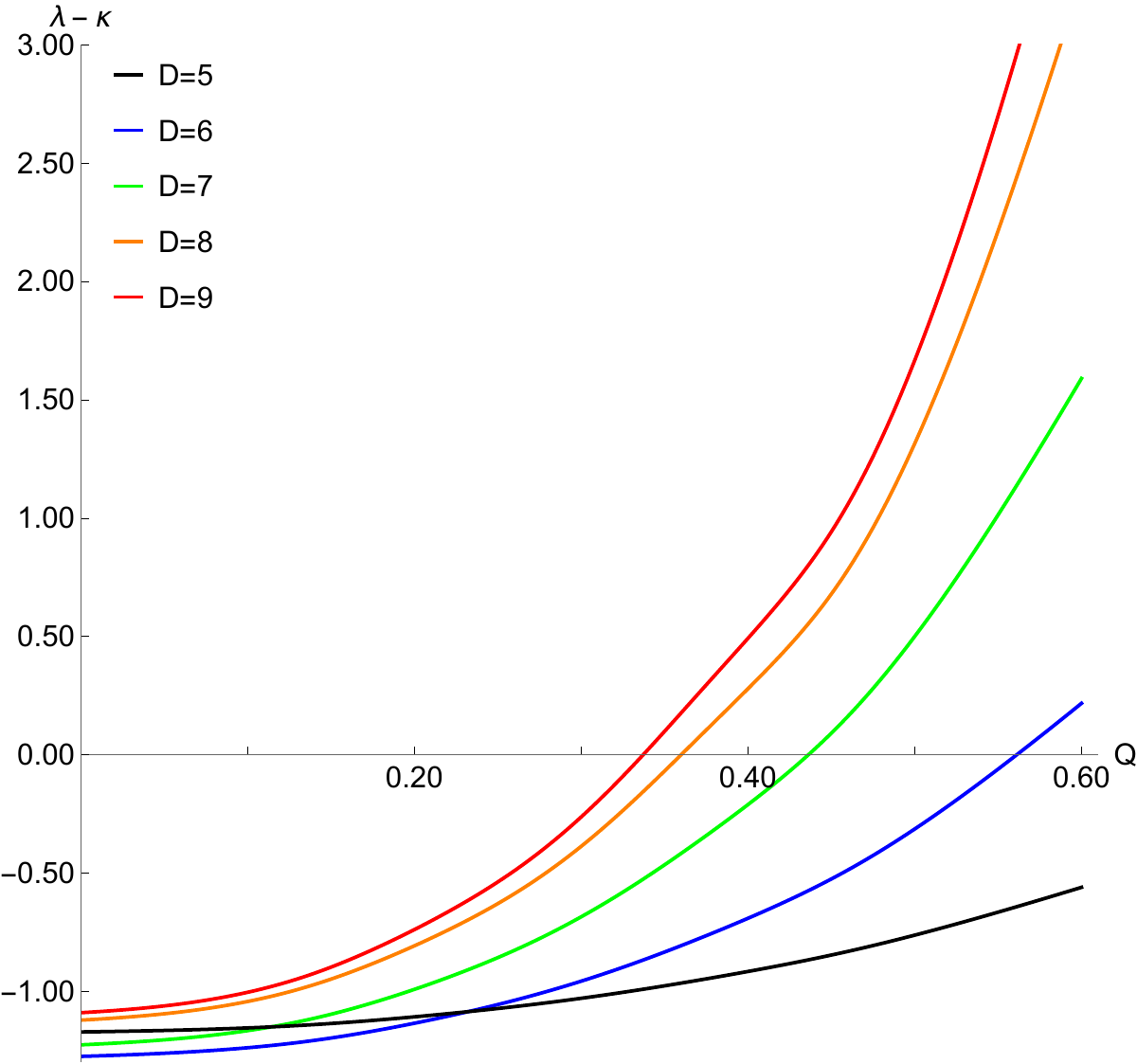}
		\subcaption{$\alpha=0.005$, $S=0.12$.}
		\label{f52}
	\end{minipage}
	\caption{Violation of the chaos bound as a function of the black hole charge. Panel (a) depicts the five-dimensional case, while panel (b) presents the results for spacetimes of various dimensions.}
	\label{f5}
\end{figure}

\begin{figure}[h]
	\centering
	\begin{minipage}[t]{0.48\textwidth}
		\centering
		\includegraphics[width=7cm,height=6cm]{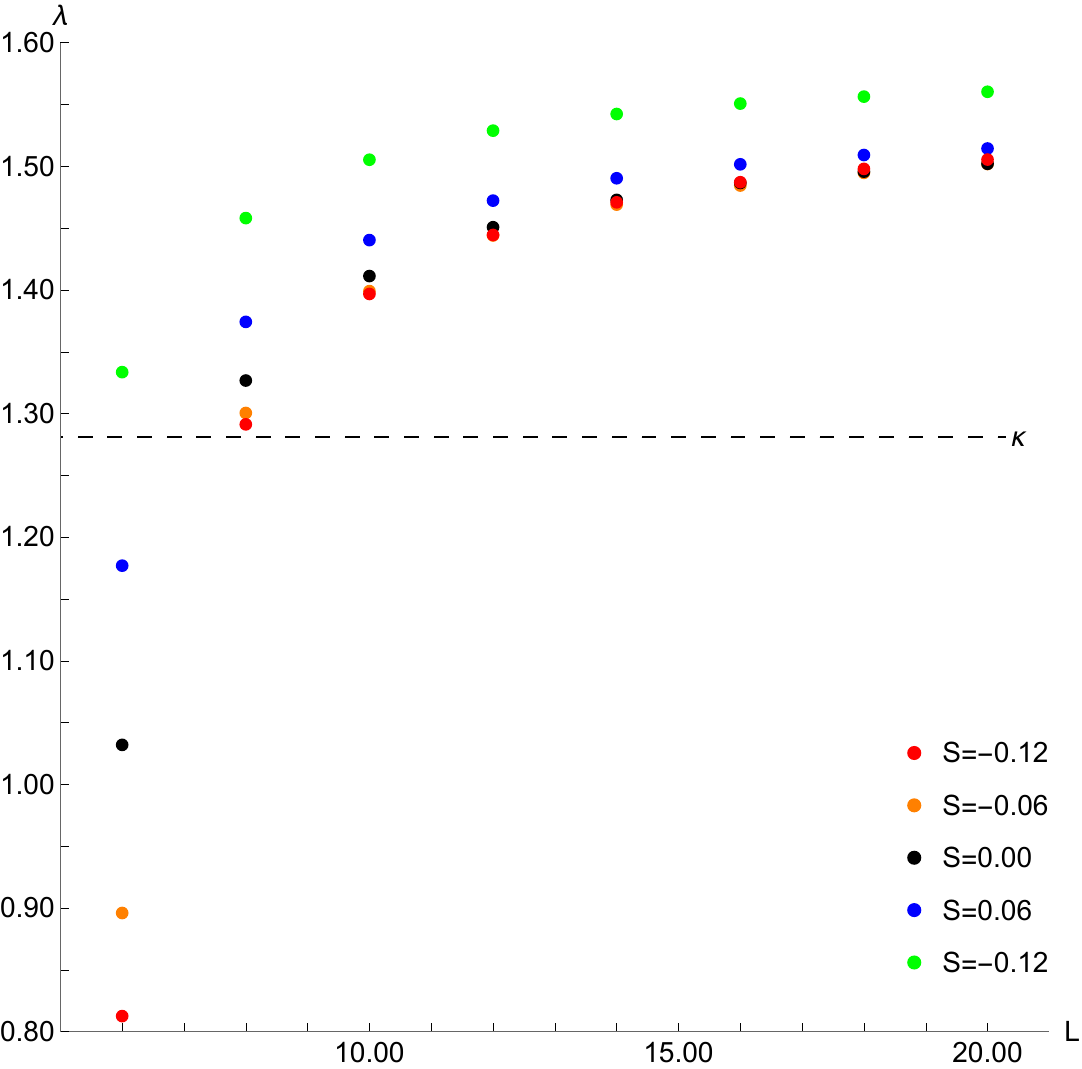}
		\subcaption{$Q=0.70$, $\alpha=0.04$, $d=5$.}
		\label{f61}
	\end{minipage}
	\begin{minipage}[t]{0.48\textwidth}
		\centering
		\includegraphics[width=7cm,height=6cm]{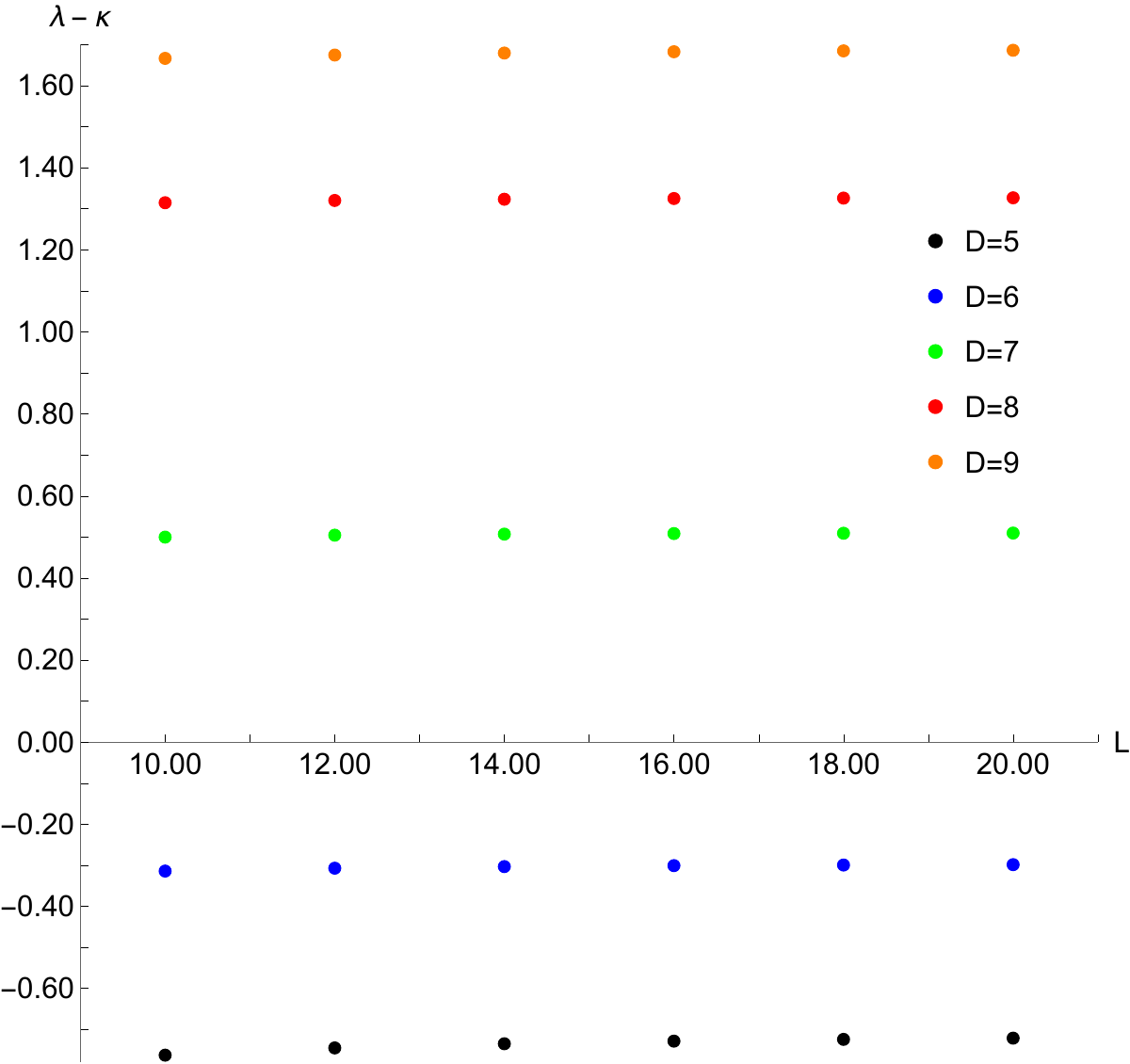}
		\subcaption{$Q=0.50$, $\alpha=0.015$, $S=0.12$.}
		\label{f62}
	\end{minipage}
	\caption{Violation of the chaos bound as a function of the particle total angular momentum. Panel (a) depicts the five-dimensional case, while panel (b) presents the results for spacetimes of various dimensions.}
	\label{f6}
\end{figure}

\begin{figure}[h]
	\centering
	\begin{minipage}[t]{0.48\textwidth}
		\centering
		\includegraphics[width=7cm,height=6cm]{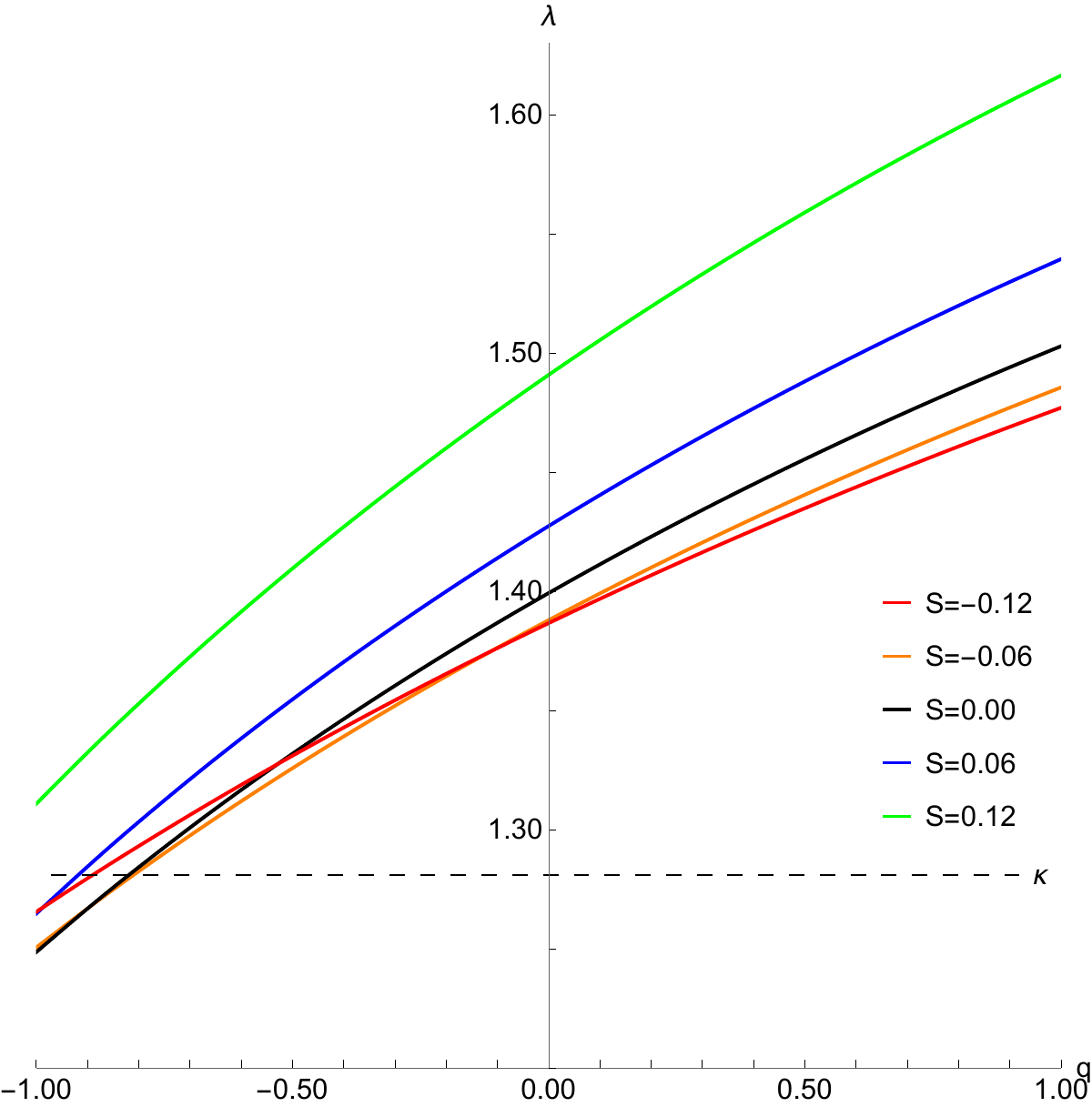}
		\subcaption{$Q=0.70$, $\alpha=0.04$, $d=5$.}
		\label{f71}
	\end{minipage}
	\begin{minipage}[t]{0.48\textwidth}
		\centering
		\includegraphics[width=7cm,height=6cm]{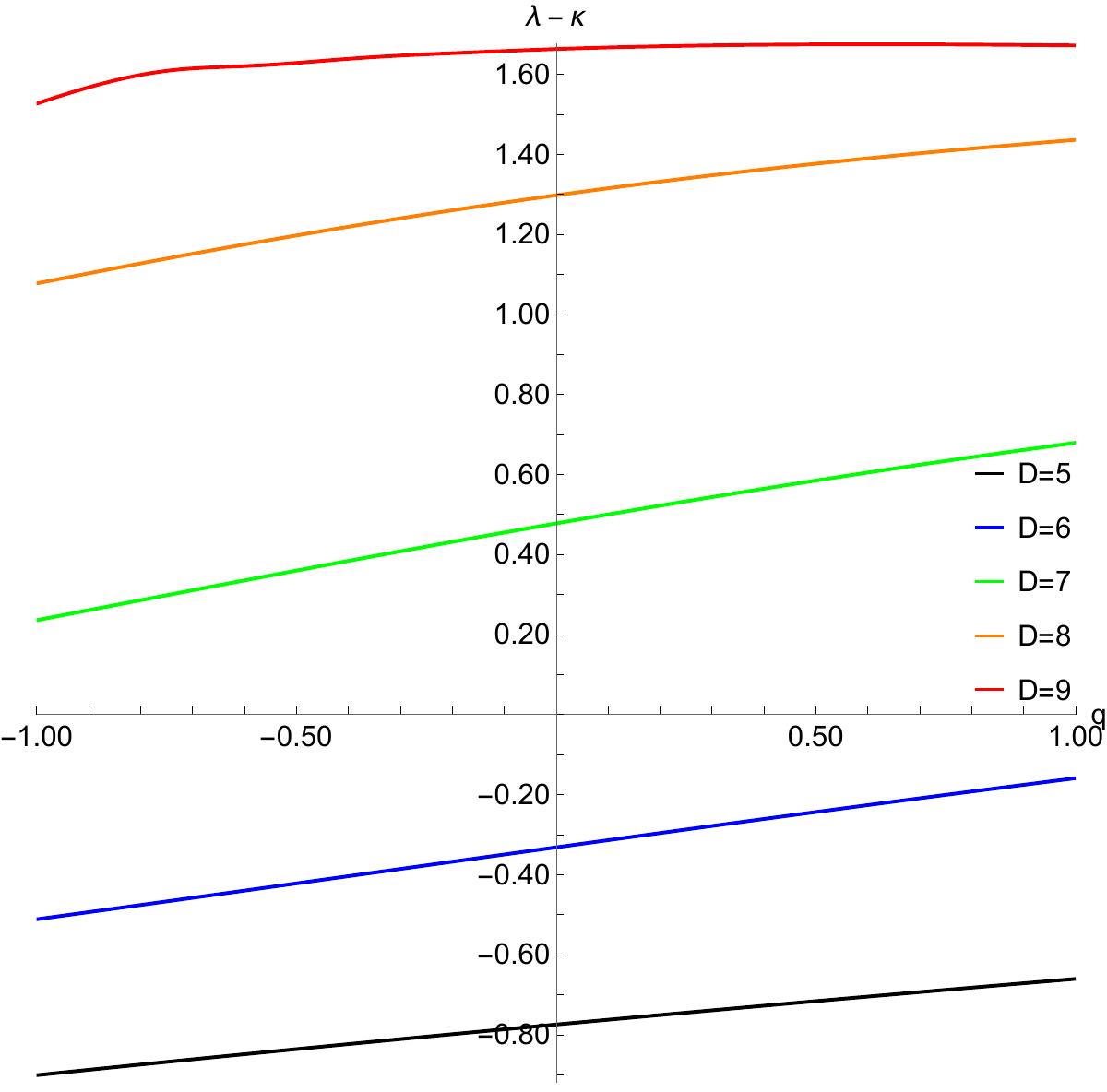}
		\subcaption{$Q=0.50$, $\alpha=0.005$, $S=0.12$.}
		\label{f72}
	\end{minipage}
		\caption{Violation of the chaos bound as a function of the particle charge. Panel (a) depicts the five-dimensional case, while panel (b) presents the results for spacetimes of various dimensions.}
	\label{f7}
\end{figure}

In this section, we numerically evaluate the surface gravity and the LE using Eqs. \eqref{eq2.1.5} and \eqref{eq2.3.3}, and present the results in Figs. \ref{f2}-\ref{f7}, in order to test the chaos bound for spinning test particles in Gauss-Bonnet-AdS spacetime. Throughout the computation, unless otherwise specified, we set $l^2=1.00$, $q=0.10$ and $L=10.00$.

Figure \ref{f2} illustrates the influence of the particle spin on the violation of the chaos bound. Figure \ref{f21} corresponds to the five-dimensional Gauss-Bonnet-AdS background. All LEs increase monotonically with the spin angular momentum in the positive direction. When the total angular momentum of the particle is greater than or equal to $8.00$, all exponents exceed the surface gravity. However, when the total angular momentum takes the value $6.00$, the exponent surpasses the surface gravity only after the particle spin exceeds a certain threshold, thereby triggering a violation of the chaos bound. Compared with other angular momentum cases, the growth rate of the  exponent reaches its maximum under this condition. The reason is that the spin angular momentum is relatively small compared with the total angular momentum, so the latter plays the dominant role in the violation, while the spin fine-tunes the effective potential through the spin-curvature coupling, thereby indirectly affecting the magnitude of the exponent. To reveal the influence of spin on the chaos bound in different spacetime dimensions, we fix the total angular momentum at $10.00$ and present Figure \ref{f22}. In this figure, to ensure the existence of unstable equilibrium orbits in all dimensionalities, the Gauss-Bonnet parameter and the black hole charge are set to smaller values than in the five-dimensional case. The results show that only in five and six dimensions do the exponents satisfy the chaos bound; all higher dimensions exhibit a violation, and in these dimensions the exponents increase monotonically with the spin angular momentum in the positive direction. In eight and nine dimensions, however, the exponents display non-monotonic behavior, first decreasing and then increasing as the spin angular momentum grows in the positive direction. Compared with Figure \ref{f21}, the variation of the exponent in five dimensions becomes milder due to the reduced Gauss-Bonnet parameter, while in seven, eight, and nine dimensions the growth rate of the exponent accelerates significantly in the region of positive spin angular momentum. The origin of these differences is that the Gauss-Bonnet term contributes dynamical effects only in five dimensions and above, whereas four dimensions correspond to pure Einstein gravity. In five and six dimensions, the Gauss-Bonnet term modifies the near-horizon spacetime geometry, allowing the chaos bound to hold. In higher dimensions, the tensorial coupling between the spin and the spacetime curvature exhibits nonlinear features, leading to the non-monotonic response observed in eight and nine dimensions. The reduction of the Gauss-Bonnet parameter weakens the coupling strength in five dimensions, while the steeper curvature gradients in higher dimensions enhance the sensitivity to spin effects, thereby accelerating the growth of the exponent.

Figure \ref{f3} displays the effect of the Gauss-Bonnet parameter on the violation of the chaos bound. In five dimensions (Figure \ref{f31}), for all spin values except $S=0.12$, the difference between the  exponent and the surface gravity first increases and then decreases as the Gauss-Bonnet parameter grows. When the this parameter reaches a certain threshold, this difference turns from negative to positive, leading to a violation. To further reveal the influence of this parameter on the chaos bound in different spacetime dimensions, we fix the particle spin at $S=0.12$ and present Figure \ref{f32}. To ensure the simultaneous existence of unstable equilibrium orbits in all higher dimensions, this parameter values chosen here are again smaller than in the five-dimensional case. The results indicate that in nine dimensions the difference between the  exponent and the surface gravity first increases and then decreases with this parameter, whereas in all other dimensions this difference increases monotonically. Moreover, for the same Gauss-Bonnet parameter, the higher the spacetime dimension, the more easily the chaos bound is violated. For instance, the exponents in eight and nine dimensions all violate the chaos bound, while those in five and six dimensions all satisfy it. The essence of this phenomenon lies in the nonlinear modification of the spacetime curvature by the Gauss-Bonnet parameter and its coupling with the dimensionality. This parameter alters the steepness of the effective potential by modifying the near-horizon geometry, thereby affecting the exponent. In higher dimensions, gravity is diluted with increasing dimensionality, and the curvature gradients become steeper, making the spin-curvature coupling more sensitive for the same Gauss-Bonnet parameter and rendering the chaos bound more easily violated. The non-monotonic behavior in nine dimensions originates from the enhanced nonlinearity of the tensorial coupling in high dimensions. The Gauss-Bonnet parameter and the dimensionality jointly regulate the rigidity of the spacetime geometry, determining whether the chaos bound holds.

Figure \ref{f4} illustrates the effect of the cosmological constant on the violation of the chaos bound. Herein, a variation in the cosmological constant induces a corresponding change in the AdS radius of the spacetime, and therefore the AdS radius can no longer be held fixed at unity. In five dimensions (Figure \ref{f41}), all the differences between the LE and the surface gravity increase with the absolute value of the cosmological constant, and all exponents exceed the surface gravity. For the same cosmological constant, the difference for the particle with spin $S=0.12$ exhibits the largest variation rate with the cosmological constant. To reveal the influence of the cosmological constant on the chaos bound in different spacetime dimensions, we fix the spin at $S=0.12$ and present Figure \ref{f42}. Owing to the reduced Gauss-Bonnet parameter and black hole charge, the trend of the difference in five dimensions is reversed compared with Figure \ref{f41}, and all exponents satisfy the chaos bound. In all other dimensions, the exponents exceed the surface gravity. In six and eight dimensions, the difference increases with the absolute value of the cosmological constant. In seven dimensions, the difference first increases and then decreases. In nine dimensions, the difference first rises rapidly, then slowly declines to an extremum, and subsequently increases slowly again.

The reason of these phenomena is as follows. A larger absolute value of the cosmological constant implies a deeper  potential well in AdS spacetime, confining the particle motion to a smaller region and intensifying the orbital chaos, which generally leads to an increase in the  exponent. The coupling between the spinning particle and the spacetime curvature makes the high-spin ($S=0.12$) particle more sensitive to the spacetime bending, therefore, its intensity of chaos exhibits the largest variation rate with the cosmological constant. The reversal of the five-dimensional behavior in Figure \ref{f42} arises because the reduction of the Gauss-Bonnet parameter and the charge alters the black hole background structure, causing the surface gravity to increase more rapidly than the  exponent, so that the chaos bound is respected. In higher dimensions, gravity decays more rapidly, and the confining effect of the potential well becomes more localized, leading to the monotonic increase in six and eight dimensions. The non-monotonic behavior in seven and nine dimensions originates from the unique topological properties of odd-dimensional gravitational theories, where multiple effects compete on different scales, forming a complex dynamical phase diagram.

Figure \ref{f5} displays the effect of the black hole charge on the violation of the chaos bound. In five dimensions (Figure \ref{f51}), for all spin values, the difference between the exponent and the surface gravity increases with the positive black hole charge. When the charge exceeds a certain threshold, the exponent surpasses the surface gravity, and a violation of the chaos bound ensues. A larger black hole charge exerts a stronger electromagnetic force on the particle, making the chaos bound more easily violated. As the charge increases, the exponent for the particle with spin $S=0.12$ is the first to violate the chaos bound, while that for $S=-0.12$ is the last. Fixing the particle spin at $S=0.12$, we examine the situation in other spacetime dimensions, as shown in Figure \ref{f52}. Although the difference between the exponent and the surface gravity increases with the positive black hole charge in all dimensions, the five-dimensional case now exhibits no violation due to the reduced Gauss-Bonnet parameter. In all other dimensions, the charge threshold for violation is lower than in the five-dimensional case. The reason is that the reduced Gauss-Bonnet parameter weakens the curvature correction in five dimensions, flattening the effective potential and preserving the chaos bound; in higher dimensions, the dilution of gravity enhances the relative strength of the electromagnetic interaction, thus lowering the violation threshold.

Figure \ref{f6} shows the influence of the total angular momentum of the particle on the  exponent. In five dimensions (Figure \ref{f61}), similar to the case of generic spherically symmetric spacetimes, all exponents increase with the total angular momentum. When the angular momentum exceeds a certain threshold, a violation occurs. In Figure \ref{f62}, owing to the reduced Gauss-Bonnet parameter and black hole charge, the exponents in five and six dimensions all satisfy the chaos bound, while those in the other dimensions all exceed it. All exponents still increase with the total angular momentum, but the growth rates are very slow. This indicates that, under this parameter setting, the influence of the spacetime dimension dominates over that of the total angular momentum. The reason is that the reduction of the Gauss-Bonnet parameter and the black hole charge reshapes the spacetime background, relatively enhancing the gravitational binding in five and six dimensions, so that the increase in the surface gravity offsets the chaotic effect induced by the angular momentum, and the chaos bound is respected. In higher dimensions, the dilution of gravity is significant, and the spacetime dimension itself becomes the dominant factor determining the chaotic behavior; hence, although the angular momentum can raise the exponent, the growth rate is very slow. Therefore, in the framework of modified gravity, the spacetime dimension and the Gauss-Bonnet parameter are no longer passive dynamical backgrounds but act as active regulatory factors.

Figure \ref{f7} displays the effect of the particle charge on the violation of the chaos bound. From Figure \ref{f71}, it can be observed that all exponents increase with the positive particle charge. When the particle charge exceeds a certain threshold, the exponent surpasses the surface gravity, triggering a violation. Notably, even when the particle charge is zero, all the exponents already exceed the surface gravity. This indicates that, under this parameter setting, the intensity of chaos in the system has already breached the chaos bound even without the particle charge, and the introduction of the charge further exacerbates the violation. This phenomenon suggests that the instability provided by the black hole background spacetime itself (e.g., the cosmological constant, the Gauss-Bonnet parameter, etc.) and/or the total angular momentum of the particle is already sufficient to drive the chaos beyond the surface gravity bound, and the electromagnetic force on the particle further amplifies the chaotic effect. From Figure \ref{f72}, it can be observed that the difference between the Lyapunov exponent and the surface gravity increases with the positive particle charge in all spacetime dimensions. In this case, the exponents in five and six dimensions all satisfy the chaos bound, while those in seven, eight, and nine dimensions all violate it.

\section{Conclusion and discussion}\label{sec4}

In this work, we investigated the violation of the chaos bound for spinning charged particles in Gauss-Bonnet-AdS spacetime, focusing on how particle and spacetime parameters influence the violation. The particle spin shifts the extremum of the effective potential through spin-curvature coupling, thereby altering the parameter threshold for bound violation. In five dimensions, the LEs grows monotonically with increasing the spin, whereas in eight and nine dimensions they exhibit non-monotonic behavior-first decreasing, then increasing-due to the nonlinear nature of higher-dimensional tensor couplings. The Gauss-Bonnet parameter regulates the chaos bound by modifying the near-horizon geometry. In five dimensions, the difference between the  exponent and the surface gravity varies non-monotonically with the Gauss-Bonnet parameter; in higher dimensions, gravitational dilution steepens the curvature gradient, making the bound easier to violate. The cosmological constant acts as a potential well: a larger magnitude yields stronger chaotic behavior. In odd dimensions, the non-monotonic behavior reflects the topological features of the gravitational theory. Both the black hole charge and the particle charge promote bound violation-stronger electromagnetic forces lower the violation threshold. Increasing the total angular momentum of the particle also enhances chaos, but when the Gauss-Bonnet parameter and black hole charge are small, the spacetime dimension overtakes angular momentum as the dominant factor. These results indicate that, within modified gravity, the spacetime dimension and the Gauss-Bonnet parameter are not passive backgrounds but active regulators of the chaos bound.

The introduction of the Gauss-Bonnet parameter incorporates higher-curvature corrections, modifying the asymptotic structure of the originally spherically symmetric spacetime and the shape of the effective potential. This, in turn, non-trivially modulates the parameter window for violation of the chaos bound: A stronger Gauss-Bonnet coupling makes the bound harder to suppress. As the spacetime dimension increases, the decay behavior of the gravitational and electromagnetic forces changes, causing the threshold condition for bound violation to exhibit qualitatively different features from the four-dimensional RN case. These results highlight the significant roles played by higher-curvature terms and extra dimensions in the physics of the chaos bound. The present work adopts the test-particle approximation, neglecting backreaction on the background spacetime, and considers only a limited range of spin values. Future work may incorporate these factors and further explore the influence of quantum corrections or semiclassical effects.

\end{document}